\documentstyle[12pt]{article}
\begin{document}
\newcommand{\beq}{\begin{equation}}
\newcommand{\eeq}{\end{equation}}
\newcommand{\bea}{\begin{eqnarray}}
\newcommand{\eea}{\end{eqnarray}}
\newcommand{\np}{\nabla_+}
\newcommand{\nm}{\nabla_-}
\begin{titlepage}
\begin{center}
            \hfill       CERN-TH.6726/92\\
            \hfill       NYU-TH-92/11/02\\
            \hfill       UCLA-92/TEP/39
\end{center}
\vskip .3in
\begin{center}
{\large \bf NON LINEAR REALIZATIONS OF $SU(2)\times U(1)$ IN THE
MSSM:\\
 MODEL INDEPENDENT ANALYSIS AND $g-2$ OF $W$
BOSONS}
\end{center}
\vskip .2in
\begin{center}
{\bf Sergio Ferrara}\footnotemark\footnotetext{Work supported in part by
the
Department of Energy of the United States under contract no.
DOE-AT03-88ER40384
Task C.}, {\bf Antonio Masiero}\footnotemark\footnotetext{On leave of
absence from
Istituto Nazionale di Fisica Nucleare, Sezione di Padova, Padova, Italy}
\vskip .1in
{\it Theory Division, CERN, Geneva, Switzerland}
\\  and \\ {\bf Massimo Porrati}\footnotemark\footnotetext{On leave of
absence
from Istituto Nazionale di Fisica Nucleare, Sezione di Pisa, Pisa,
Italy}
\vskip .1in
{\it Department of Physics, NYU, New York, NY 10003 USA}
\vskip .4in
 \bf{ABSTRACT}\end{center}
\begin{quotation}  \noindent
We perform a model-independent analysis of the spontaneously broken
phase of
an $SU(2)\times U(1)$ supersymmetric gauge theory, by using a non-linear
parametrization of the Goldstone sector of the theory. The non-linear
variables
correspond to an $SL(2,C)$ superfield matrix in terms of which a
non-linear
Lagrangian can be constructed, and the pattern of supersymmetry breaking
investigated. The supersymmetric order parameter is the V.E.V. of the
neutral
pseudo-Goldstone boson. Some applications of this technique are
considered, in
relation to the minimal supersymmetric standard model, and to determine
the
$g-2$ of the $W$-bosons in the limit of large top mass.
\end{quotation}
\vfill
CERN-TH.6726/92 \hfill \\
November 1992 \hfill
\end{titlepage}
\vfill
\eject
\section{Preliminaries}
One of the main candidate theory for an extension of the Standard Model
incorporating new physics is its supersymmetric version. It is based on
supersymmetric Yang-Mills theory~\cite{1}
with gauge group $SU(3)\times SU(2)\times
U(1)$ spontaneously broken to $SU(3)\times U(1)_{em}$ plus chiral and
vector
multiplets of N=1 supersymmetry\cite{2}.

If supersymmetry is broken by soft terms~\cite{3},
then the model retains the good
ultraviolet properties of supersymmetric theories~\cite{4},
and gives a realistic
candidate for the solution of the hierarchy problem~\cite{5},
in agreement with present
experimental constraints, and predicting a wealth of new
physics in the TeV range.

Soft breaking terms may be thought of as relics of spontaneously broken
supergravity~\cite{6},
which, in turns, may arise as the low-energy effective action
of superstring theory~\cite{7}.

In the present letter we would like to give a model independent analysis
of
the spontaneously broken $SU(2)\times U(1)$ gauge theory, in order to
show
that many of the predictions encoded in the Minimal Supersymmetric
Standard
Model (MSSM)~\cite{8}
are actually universal, and to show how different supersymmetric
extensions of the Standard Model, with or without broken SUSY, can be
classified in this contest.

As a partial application of our analysis, we will obtain the MSSM, and
derive
non-minimal versions of it, with softly broken supersymmetry, as well as
model with unbroken SUSY. In the last case we will compute the $g-2$ of
$W$ vector bosons, verifying, in this particular case, a general sum
rule
of magnetic dipole moments of arbitrary electrically charged
supersymmetric
particles~\cite{9}.
\section{Non-Linear Realizations of $SU(2)\times U(1)$}
We want to describe the spontaneously broken phase of an arbitrary
$SU(2)\times U(1)$ gauge theory plus matter,
with a residual unbroken gauge symmetry $U(1)_{em}=T_3+Y/2$.

The model independent part of this theory is described by a non-linear
K\"ahlerian sigma model~\cite{10},
constructed with the coset representatives
of $SL(2,C)\times  GL(1,C)/GL(1,C)_{em}$.
The coordinates of this manifold are the three Goldstone bosons of
$SU(2)\times U(1)/U(1)_{em}$ and their SUSY partners, which are
pseudo-Goldstone particles. The first three are the longitudinal degrees
of
freedom of the $W$ and $Z$ particles, the others complete massive vector
multiplets, in case of unbroken SUSY.
An element of this coset is described by the two-by-two holomorphic
matrix
\beq
U(\xi)=e^{i\xi\cdot\sigma/2}, \;\;\; \det U=1.
\label{1}
\eeq
It transforms as follows under $SL(2,C)\times GL(1,C)$
\beq
U\rightarrow e^{+i\Lambda} U e^{-1/2 i\Sigma \sigma_3},\;\;\;
\Lambda\equiv
{1\over 2}\Lambda_i\sigma_i.
\label{2}
\eeq

The minimal
sigma model Lagrangian is a K\"ahlerian sigma model with K\"ahler
potential
\beq
K(\xi,\bar{\xi})={\rm Tr}\, (U^\dagger U)={\rm Tr}\, (UU^\dagger) .
\label{3}
\eeq
This Lagrangian is invariant under $SU(2)\times U(1)$ global
transformations,
which correspond to $\Lambda$, $\Sigma$ real. It is also invariant under
arbitrary local $\Lambda$, $\Sigma$ transformations, provided we
introduce
gauge fields $W=gW_i\sigma_i/2$, $B=g'Y\sigma_3/2$ transforming as
\beq
e^W\rightarrow e^{+i\Lambda^\dagger} e^W e^{-i\Lambda },
\label{4}
\eeq
\beq
e^B\rightarrow e^{i/2\Sigma\sigma_3} e^B e^{-i/2\Sigma^\dagger\sigma_3}.
\label{5}
\eeq
We may also describe non-minimal couplings corresponding to the most
general
$SU(2)\times U(1)$-invariant K\"ahler potential, constructed in terms of
$U$ and $U^\dagger$. This K\"ahler potential is only function of two
$SU(2)\times U(1)$-invariant variables $K(A,C)$, with $A={\rm
Tr}\,U^\dagger U$,
$C={\rm Tr}\,U^\dagger U\sigma_3$.The minimal Lagrangian, discussed in
this
paper, corresponds to $K(A,C)=A$.

Let us define a new composite superfield
\beq
\hat{U} = e^W U e^B,\;\;\;\overline{D}_{\dot\alpha} U = 0 .
\label{6}
\eeq
The modified (gauged) Lagrangian is
\beq
\mu^{-2}{\cal L} (U,W,B)= \left[{\rm Tr}\,
(U^\dagger \hat{U})\right]_D = \left[{\rm Tr}\, (U^\dagger e^W U e^B)
\right]_D = \left[{\rm Tr}\, (\hat{U} U^\dagger)\right]_D,
\label{7}
\eeq
where $\mu$ gives the scale of $SU(2)\times U(1)$ breaking\footnote{Note
that
$\xi_i = {\phi_i\over \mu}$, in terms of dimensions 1 scalar fields
$\phi_i$.}
{}.
It is convenient to define the two (matrix) superfields \beq
V_R=U^\dagger \hat{U}
\label{8}
\eeq
transforming as $V_R\rightarrow e^{i/2\Sigma^\dagger\sigma_3}V_R
e^{-i/2\Sigma^\dagger\sigma_3}$, and
\beq
V_L=\hat{U} U^\dagger,
\label{9}
\eeq
transforming as $V_L\rightarrow e^{+i\Lambda^\dagger}V_L
e^{-i\Lambda^\dagger}$.
By means of these fields one can define the $SU(2)\times U(1)$ gauge
invariant
superfields
\bea
Z^0 &=& {\rm Tr}\,(V_R\sigma_3), \label{10} \\
W^\pm  &=& {\rm Tr}\,(V_R\sigma^\pm), \label{11}
\eea
whose $\theta \sigma \bar{\theta}$ components are
the $Z_\mu^0$, $W_\mu^\pm$ particles up to sigma model corrections.
The first is $SU(2)\times U(1)$ invariant and the second transforms,
under
this group, as
\beq
W^\pm \rightarrow e^{\pm i\Sigma^\dagger}W^{\pm}.
\label{12}
\eeq
The complex conjugate superfields $\overline{W}^\pm$ transform as
\beq
\overline{W}^\pm \rightarrow e^{\mp i\Sigma}\overline{W}^\pm, \;\;\;
 \overline{W}^\pm = W^{\mp}e^{\mp g^\prime Y}.
\label{13}
\eeq
The $SU(2)\times U(1)$ gauge invariant expression for the $g-2$ of the
$W$
supermultiplet is
\beq
\left[ W^+ \buildrel \leftrightarrow \over D_\alpha W^- V^\alpha
\right]_D,
\label{14}
\eeq
where $V_\alpha= \bar{D}\bar{D} D_\alpha V$ is the e.m.
superfield-strength.

The invariant mass terms are simply given by
\beq
\left[ W^+W^-\right]_D,\;\;\; \left[(Z^0)^2\right]_D .
\label{15}
\eeq
Instead, the invariant quartic couplings are given by
\bea
& &
\left[ W^+W^- W_{\mu}^+W^{\mu\, -}\right]_D, \;\; \left[(Z^0)^2Z_{\mu}^0
Z^{\mu 0} \right]_D,\;\; \left[ W^+W^- Z_{\mu}^0 Z^{\mu\, 0}\right]_D ,
\nonumber \\
& & \left[W^+W^+\overline{W}_\mu^+\overline{W}_\mu^+\right]_D,\;\;\;
\left[W^-W^-\overline{W}_\mu^-\overline{W}_\mu^-\right]_D
\label{16}
\eea
where
\beq Z^0_{\mu} =\sigma^{\alpha\dot{\alpha}}_\mu
[\bar{D}_{\dot{\alpha}},D_\alpha]Z^0,\;\;\;
W^\pm_\mu = \sigma^{\alpha\dot{\alpha}}_\mu
[\bar{D}_{\dot{\alpha}},e^{\pm g^\prime Y} D_\alpha]W^\pm .
\label{17}
\eeq
The K\"ahler potential of the sigma model Lagrangian reads
\bea
K(\xi,\bar{\xi})&=&{\rm Tr}\,
(e^{-i/2 \bar{\xi}\cdot\sigma}e^{i/2\xi\cdot\sigma})=
\nonumber \\
&=& 2\cos {1\over 2} \sqrt{\xi_i^2}\cos {1\over 2} \sqrt{\bar{\xi}_i^2}
2{\xi_i\bar{\xi}_i\over \sqrt{\xi_i^2}\sqrt{\bar{\xi}_i^2}} \sin{1\over
2}\sqrt{\xi_i^2} \sin {1\over 2}\sqrt{\bar{\xi}_i^2}.
\label{18}
\eea
It is easy to see that this Lagrangian reduces to the non-linear
Standard
Model Lagrangian~\cite{11}
if we set to zero the pseudo-Goldstone bosons.

It is now straightforward to discuss the features of this Lagrangian,
even in
the presence of soft supersymmetry breaking~\cite{3}.
The most general softly broken
Lagrangian\footnote{By soft breaking terms we mean, in this context,
terms which
are bilinear in the $U$ matrix.}
 which preserves $SU(2)\times U(1)$ is given by
\beq
{\cal L}_{YM} + \mu^4{\rm Tr}\,
\left[ U^\dagger \hat{U}\right]_D - \mu^2 {a \over 4}{\rm Tr}\, \left.
(U^\dagger
U)\right|_{ \rm first\; component} - \left. \mu^2 {b \over 4}{\rm Tr}\,
(U^\dagger U\sigma_3) \right|_{ \rm first\; component} .
\label{19}
\eeq
The D terms produce a quartic term in the potential of the form
\beq
\mu^4 {g^2\over 8} {\rm Tr}\,(U^\dagger \sigma_i U)^2 +
\mu^4{g'^2\over 8}[{\rm
Tr}\,(U^\dagger U\sigma_3)]^2.
\label{20}
\eeq
Notice that the $b$ parameter can be viewed as a Fayet-Iliopoulos
term~\cite{12}, while
the $a$ parameter is a true soft breaking term.
The $a$ parameter could become a supersymmetric term in two ways. Either
as a
Fayet-Iliopoulos term with respect to a new $U(1)$ gauge
group~\cite{13},
under which
the $U$ field has a definite nonzero charge, or as a relic of the
supergravity
coupling in the limit $M_{Planck}\rightarrow \infty$,
$m_{3/2}=$constant~\cite{6}.
Only the second possibility seems to give rise to a phenomenologically
acceptable version of MSSM.

Because of gauge invariance we may set ${\rm Re}\,\xi_i=0$, thus, we
have a
Lagrangian depending only on the following parameters:
the complex variable
$\xi_+={\rm Im}\,\xi_1 +i{\rm Im}\,\xi_2$, and the real variable
$\xi={\rm
Im}\,\xi_3$.

This Lagrangian has an absolute ``minimum'' at $\xi_\pm=0$, and $\xi$
given by
the universal relation
\beq
2\mu^2(g^2+g'^2)= -{a\over \cosh \xi } + {b\over \sinh \xi}.
\label{21}
\eeq
We may define $e^\xi =\tan \beta$ $(0\leq \beta \leq \pi/2)$, then we
have
\bea
\sin 2\beta &=&
{2{\rm tan} \beta \over 1 + ({\rm tan}\beta)^2}={1\over \cosh \xi},
\nonumber \\
\tan 2\beta &=& -{1\over \sinh \xi}.
\label{22}
\eea
Therefore, in the more conventional variable $\beta$, eq.~(\ref{21})
reads~\cite{14}
\beq
2\mu^2(g^2 +g'^2) = -a\sin 2\beta -b\tan 2\beta,
\;\;\; 0\leq 2\beta \leq \pi .
\label{23}
\eeq
The case $a=b=0$ corresponds to unbroken supersymmetry~\cite{2},
the $D_3$ and $D_Y$
terms, proportional to $\sinh \xi$ then vanish at $\xi=0$. The case
$a=0$,
$b\neq 0$ corresponds to spontaneously broken supersymmetry~\cite{12}.
The case $a\neq 0$, $b\neq 0$ corresponds to softly broken
supersymmetry.

We notice that in this analysis there is only a charged Higgs $H^\pm$,
and a
neutral Higgs $H^0$, which are the superpartners of the $W^\pm$ and
$Z^0$
particles. This is the minimal set of states which must exist in any
supersymmetric model, independently of whether $SU(2)\times U(1)$ is
a weakly coupled theory or a strongly coupled one (with dynamical
breaking of
gauge symmetry).
\section{Connection with the MSSM}
The MSSM is obtained by introducing a $GL(2,C)$ matrix $\Phi$,
instead of a $SL(2,C)$ matrix $U$ as follows
\beq
\Phi = SU=Se^{i/2\xi\cdot \sigma},\;\;\; \det \Phi =S^2,
\label{24}
\eeq
where $S$ is a chiral superfield. The $\Phi$ transformation is identical
to
the $U$ transformation, under $SU(2)\times U(1)$, but now $\Phi$ admits
linear realization in terms of two doublet chiral superfields $H_{(1)}$,
$H_{(2)}$, which transform
linearly under $SU(2)\times U(1)$ as follows
\footnote{Note that this parametrization excludes the points for which
det
$\Phi=0$.}
 \beq
\Phi =
\left( \begin{array}{cc} H_{(1)1} & H_{(2)1} \\ H_{(1)2} & H_{(2)2}
\end{array}
\right) ,\;\; H_{(1)}\rightarrow e^{+i\Lambda} e^{-i/2\Sigma}
H_{(1)},\;\;
H_{(2)} \rightarrow e^{+i\Lambda} e^{+i/2\Sigma} H_{(2)} .
\label{25}
\eeq
It is immediate to see that $S^2=H_{(1)}\cdot H_{(2)}$, and that
\bea
{H_{(1)1} + H_{(2)2} \over 2} = S\cos {1\over 2} \sqrt{\xi_i\xi_i}, & &
H_{(2)1}=iS{\xi_+\over \sqrt{\xi_i\xi_i}}
\sin {1\over 2}\sqrt{\xi_i\xi_i}, \nonumber\\
{H_{(1)1} - H_{(2)2} \over 2} = iS{\xi_3 \over \sqrt{\xi_i\xi_i}}
\sin {1\over 2} \sqrt{\xi_i\xi_i}, & &
H_{(1)2}=iS{\xi_-\over \sqrt{\xi_i\xi_i}}
\sin {1\over 2} \sqrt{\xi_i\xi_i}.
\label{26}
\eea
Therefore, the non linear model is obtained from the linear one by
integrating
out the $S$ fields

Note that if we set
the pseudo-Goldstone to zero, ${\rm Im}\, \xi =0 $,
and we define ${\rm Re}\, \xi \equiv \sigma $,
$H_{(2)2}=\overline{H}_{(1)1}$, $H_{(2)1}=-\overline{H}_{(1)2}$ and we
recover
the usual parametrization of the SM Higgs sector, where
$\Phi=\sigma e^{i/2 {\rm Re}\, \xi_i\sigma_i}$, $\det \Phi
=\Phi\Phi^\dagger
=\sigma^2$.

The Higgs-sector Lagrangian is now given by the same expression as in
eq.~(\ref{7})  with $U\rightarrow \Phi $ with an additional potential
for the
$S$ field
\bea
V &=& {g^2\over 8}\left[({\rm Tr}\,
\Phi^\dagger \sigma_i \Phi)^2\right]_D + {g'^2\over 8}
\left[({\rm Tr}\, \Phi^\dagger \Phi\sigma_3)^2\right]_D
+ {a \over 4}{\rm Tr}\,\Phi^\dagger \Phi +
\nonumber \\ & & {b \over 4}{\rm Tr}\, \Phi^\dagger\Phi \sigma_3 + V(S).
\label{20'}
\eea
The most general potential $V(S)$ contains a supersymmetric part
$V(S)_{SUSY}$
plus a SUSY breaking part $V(S)_{breaking}$. By setting $V(S)_{SUSY}=0$,
$V(S)_{breaking}=\left. -(m^2/2) S^2\right|_{\rm first\; component}$,
we recover the
MSSM. By choosing instead a $V(S)_{SUSY}\neq 0$, we may obtain a
non-minimal
version of the SSM, with or without supersymmetry breaking, which
corresponds
to having integrated out an additional $SU(2)\times U(1)$-singlet chiral
multiplet~\cite{15}.

Let us discuss in some detail the MSSM and a supersymmetric model
with
\beq
V(S)= \left[{\lambda \over 3} S^3 -\tau S\right]_F .
\label{fa}
\eeq
In both these
models we get a second equation which stabilizes the $S$ degree of
freedom. In
the second
model with unbroken supersymmetry, the minimum is reached when the
$D$ and $F$ terms vanish. This gives
\beq
\xi=0,\;\;\; S^2={\tau\over \lambda}.
\label{22'}
\eeq
If one defines the paramenters $v_1$, $v_2$, related to $\xi$ and $S$ by
the
general formula $e^\xi=v_2/v_1$, $ S^2=v_1v_2$, eq.~(\ref{22'}) implies
\beq
v_1=v_2,\;\;\; v_1\cdot v_2 = {\tau\over \lambda}.
\label{23'}
\eeq
In the MSSM, when the Higgs sector is parametrized by the three
constants
$a$, $b$, $m^2$, we obtain, for the $S$ field, the minimum condition
\beq
\cosh \xi = {a\over m^2},
\label{24'}
\eeq
in the range of parameters $a^2>m^4>a^2-b^2$.
Eq.~(\ref{24'}) turns out
to be the standard condition relating $\sin 2\beta$, $m_1$,
$m_2$, and $m_3$~\cite{16}, if one takes into account that
\beq
a=2(m_1^2 +m_2^2),\;\;\; b=2(m_1^2-m_2^2),\;\;\; m^2=4m_3^2.
\label{24''}
\eeq
In the case of generic non-minimal extensions of the SSM, and when the
additional degrees of freedom are integrated out, we end up with a
generic
$V(S)$ in eq.~(\ref{20'}). In this case eq.~(\ref{21}) still applies but
with
the replacements given by
\beq
\mu^2\rightarrow S\bar{S},\;\;\; a\rightarrow a+\left|{1\over
S}{\partial
W\over \partial S}\right|^2,
\label{24'''}
\eeq
where
\beq
\frac{1}{4}({\rm Tr}~\Phi^{\dag}\Phi) \left|
\frac{1}{S}~\frac{\partial W}{\partial S} \right|^2
=\left[{V(S)_{SUSY}}\right]_F\;.
 \label{24m}
 \eeq
 \section{$g-2$ of $W$-Particles
and Supersymmetric Sum Rules} Non-linear realizations of $SU(2)\times
U(1)$ give
a powerful tool to compute some physically meaningful quantities, like
the
magnetic moments of elementary particles.

Recently, it has been shown that unbroken supersymmetry relates, in a
model-independent way, the magnetic transitions between states of
different
spin within a given charged massive multiplet of arbitrary
spin~\cite{9}.

Any given massive multiplet with $J_{max}=J+1/2$ contains four charged
particles with the same charge and spin $J+1/2$, $J$, $J$, and $J-1/2$.
Let us define the gyromagnetic ratio $g_J$ of a given particle as
\beq
\vec{\mu}={e\over 2M} g_J \vec{J}.
\label{25'}
\eeq
Then, in a given multiplet with $J_{max}>1$ we have the following sum
rules
\beq
g_{J+1/2}-2=2Jh_J,\;\; g_J-2=(2J+1)h_J,\;\; g_{J-1/2}-2=(2J+2)h_J,
\label{26'}
\eeq
where $h_J$ is the magnetic transition between spin $J+1/2$ and $J-1/2$.
There
is no magnetic transition between the two spin-$J$ states. The sum rules
read
differently for $J=0$ and $J=1/2$ multiplets, which contain scalar
particles.

For $J=0$ (chiral) multiplets, the sum rule becomes $g_{1/2}=1/2$, as
shown a
long time ago in ref.~\cite{17}.
For $J=1/2$ (the vector multiplet) the sum rules
are instead
\beq
g_{1/2}-2=2h,\;\;\; g_1-2=h.
\label{27'}
\eeq
Here, $h$ is the magnetic transition between the spin-0 and spin-1
states.

In a renormalizable theory of spin-1/2 and spin-1 particles, the
tree-level
value of the gyromagnetic moment is 2~\cite{18,19}.
Arguments have been given to show that
this is a general phenomenon, which should occur in any tree-level
unitary
theory\footnotemark~\cite{18,20}.
\footnotetext{We see that for $J>0$ supermultiplets $g_J=2$ corresponds
to
vanishing transition magnetic moments $h_J$. For $J=1/2$ (i.e. for
vector
multiplets) this interaction would be power-counting non renormalizable,
and
this fact explains why $h_{1/2}=0$ at tree level in the SSM.}
However quantum effects can spoil this property. In a supersymmetric
theory,
if SUSY is unbroken, $g-2=0$ for any chiral multiplets, as implied by
our sum
rules. This applies, in particular, to quarks and leptons in the MSSM.
However, for spin-1 multiplets, $h$ could be nonzero due to loop
effects.
Indeed, this was shown to happen in an explicit calculation,
in the limit of vanishing quark and lepton masses~\cite{21}.

Eq.~(\ref{27'}) implies that
supersymmetry relates the following couplings~\cite{9,21,22}
\bea
& & {g\over M_W}(\psi^+\sigma_{\mu\nu} \lambda^- -
\psi^-\sigma_{\mu\nu} \lambda^+)F^{\mu\nu},\;\;\;
gW^+_\mu W^-_\nu F^{\mu\nu},\nonumber\\
& & {g\over M_W}\epsilon^{\mu\nu\rho\sigma}(\partial_{\mu}H^+W_\nu^-
-\partial_\mu H^-W^+_\nu)F_{\rho\sigma}.
\label{28'}
\eea
These Lagrangian couplings are responsible for the magnetic-moment
relations~(\ref{27'}) in the case of spontaneously broken $SU(2)\times
U(1)$.
To prove this statement we do not need to have a renormalizable theory,
indeed, the one-loop result will satisfy eq.~(\ref{27'}) regardless of
the
existence of the $S$ field,
which is needed in order to have a linear realization of $SU(2)\times
U(1)$. Moreover, the same relations are insensitive to the value of the
Yukawa couplings.
Indeed, we may take a physically meaningful limit, wherein all
quark and lepton masses vanish, with the exception of the top mass,
which is
sent to infinity~\cite{23}.
In this case, the effective theory gives rise to
eq.~(\ref{27'}) by a combination of one-loop graphs and the tree-level
point-interactions
produced by integrating out the top supermultiplet.
Of course, eq.~(\ref{27'}) cannot be explained through the gauge anomaly
cancellation that takes place in the renormalizable theory, as in
ref.~\cite{21}.
However, supersymmetry holds because the gauge anomalies, in the limit
of infinite top mass, are cancelled by local Wess-Zumino terms, which
are
present in the effective action~\cite{24}.
The quark and lepton contributions due to
loops of massless quarks are added to the tree-level terms, and produce
an
effective interaction as in eq.~(\ref{14}). The role of supersymmetric
Wess-Zumino terms in this analysis will be discussed elsewhere.

By the same means one can also get additional informations about the
MSSM.
One is about the possible existence of a transition magnetic moment
between
the neutralinos belonging to the the $S$ supermultiplet and the
neutralinos of
the $Z^0$  multiplet
\beq
\left[ S(\bar{D}\bar{D} D_\alpha Z^0) V^\alpha \right]_F.
\label{29'}
\eeq
A transition magnetic moment between the two neutralinos of the $Z^0$
multiplet is not allowed when supersymmetry is unbroken.
\section{Lepton-Number Violating Interactions}
In the MSSM, if lepton-number conservation is not imposed, it is
possible to
write $SU(2)\times U(1)$ gauge-invariant but lepton-number violating
interactions which couple the leptons to the Higgs sector~\cite{25}.
These interactions
may give rise to mass mixing between lepton multiplets and Higgs
multiplets,
as well as transition magnetic moments between the charged leptons and
the
charginos. Moreover, one can get  a transition magnetic moment between
the
neutrinos and a neutralino in the $Z^0$ multiplet. However,
supersymmetry
forbids  a magnetic moment between the neutrinos and the neutralinos of
the $S$
multiplet.

We give now some explicit formulas for these new interactions.
Consider the lepton doublet, transforming as
\beq
E^-\rightarrow e^{+i\Lambda} e^{-i/2\Sigma}E^-.
\label{30'}
\eeq
Under $SU(2)\times U(1)$ the quantity $E^-\sigma_2$ transforms as
\beq
E^-\sigma_2 \rightarrow E^-\sigma_2 e^{-i/2\Sigma}e^{-i\Lambda}.
\label{31'}
\eeq
Then we can construct the quantity $E^-\sigma_2U$, transforming as
\beq
E^-\sigma_2U\rightarrow E^-\sigma_2Ue^{-i/2\Sigma(I+\sigma_3)}.
\label{32'}
\eeq
Therefore the up and down components of this quantity have definite
$U_{e.m.}(1)$ charge
\bea
(E^-\sigma_2U)_{up} & \rightarrow & (E^-\sigma_2U)_{up}e^{i\Sigma},
\label{33'}\\
(E^-\sigma_2U)_{down} &\rightarrow & (E^-\sigma_2U)_{down}.
\label{34'}
\eea
The lepton-violating interaction term is therefore
\beq
\left[(E^-\sigma_2U)_{down}\right]_F
\label{35'}
\eeq
In the linear theory, this corresponds to the following MSSM interaction
\beq
E^-\sigma_2H_2.
\label{36'}
\eeq

If we take the positively charged lepton singlet $E^+$, transforming as
\beq
E^+\rightarrow E^+ e^{+i\Sigma}
\label{37}
\eeq
under $SU(2)\times U(1)$, we may form the doublet
\beq
\psi=\left(\begin{array}{c} (E^-\sigma_2U)_{up}\\ E^+
\end{array}\right),
\label{37'}
\eeq
which transforms as $\psi\rightarrow e^{-i\Sigma\sigma_3}\psi$.
The lepton-chargino transition magnetic moment is
\beq
\left[\psi(\bar{D}\bar{D} e^{+2B}D_\alpha W)V^\alpha\right]_F,
\label{38'}
\eeq
whereas the neutrino-neutralino transition magnetic moment is
\beq
\left[ (E^-\sigma_2U)_{\rm down}(\bar{D}\bar{D}D_\alpha
Z^0)V^\alpha\right]_F.
\label{39'}
\eeq
Eqs.~(\ref{38'},\ref{39'}) are manifestly $SU(2)\times U(1)$ invariant.
Notice
that all possible extensions of the interactions of the quark and lepton
sector are given only by gauge-invariant combinations of the $U$ matrix,
and
therefore are insensitive to the detailed choice of a given linear
realization
of the gauge group. A detailed analysis of different models, with
explicit
calculations of masses and couplings will be given elsewhere.

Finally, it should be pointed out that the non-linear approach is
particularly
suitable to couple the system to supergravity and to disentangle the
effects
related to the spontaneous breaking of gauge
symmetry and local supersymmetry.\vskip .1in \noindent
Acknowledgements\\
We would like to thank E. D'Hoker for enlightening discussions.

\end{document}